\documentclass[12pt]{article}
\usepackage{epsfig,epsf}
\def\beq{\begin{equation}}
\def\eeq{\end{equation}}
\def\eq#1{{Eq.~(\ref{#1})}}
\newcommand{\as}{\alpha_S}
\newcommand{\bra}[1]{\langle #1 |}
\newcommand{\ket}[1]{|#1\rangle}

\newcommand{\tr}{{\rm tr}}

\newcommand{\T} {\mbox{T}}

\begin{document}
\begin{flushright}
BNL--NT--99/8\\
TAUP 2611--99\\
hep-ph/9912216

\end{flushright}

\begin{center}
{\large\bf Scale Anomaly and ``Soft'' Pomeron in QCD}
\vskip1cm

Dmitri Kharzeev${}^{a),b)}$  and Eugene Levin${}^{b),c)}$
\vskip1cm
{\it a) RIKEN-BNL Research Center,}\\
{\it Brookhaven National Laboratory,}\\
{\it Upton, NY 11973 - 5000, USA}\\
email: kharzeev@bnl.gov 
\vskip0.3cm
{\it b) Physics Department,}\\
{\it Brookhaven National Laboratory,}\\
{\it Upton, NY 11973 - 5000, USA}\\
\vskip0.3cm
{\it c ) HEP Department, School of Physics,}\\
{\it Raymond and Beverly Sackler Faculty of Exact Science,}\\
{\it Tel Aviv University, Tel Aviv 69978, ISRAEL}\\
email: leving@post.tau.ac.il; elevin@quark.phy.bnl.gov


\end{center}
\bigskip
\begin{abstract}
We propose a new non--perturbative approach to hadronic interactions 
at high energies and small momentum transfer, 
which is based on the scale anomaly of QCD and 
emphasizes the r\^{o}le of semi--classical vacuum fields. 
We find that the hadron scattering 
amplitudes exhibit Regge behavior and evaluate the intercept $\alpha(0)$  
 of the corresponding trajectory. Both the intercept 
and the scale for the slope of the trajectory  
appear to be determined by the energy density of non--perturbative 
QCD vacuum (the gluon condensate). Numerically, we find 
$\Delta \equiv \alpha(0) - 1 = 0.08 \div 0.1$, consistent with the values ascribed 
phenomenologically to the ``soft'' 
Pomeron. For arbitrary numbers of colors $N_c$ and flavors $N_f$, 
$\Delta$ is found to be proportional to $(N_f/N_c)^2$; however, in the large $N_c$ ($N_f$ fixed)  
limit, $\Delta \sim N_c^0$.  
\end{abstract}
\vskip0.3cm

Understanding the behavior of QCD at high energies 
and small momentum transfer is still a challenging and 
unsolved problem. In the framework of perturbation theory, 
a systematic approach was developed by Balitsky, Fadin, Kuraev 
and Lipatov \cite{BFKL}, who demonstrated that the ``leading 
log'' terms in the scattering amplitude 
of type $(g^2\ ln\ s)^n$ (where $g$ is the strong 
coupling) can be re-summed, giving rise to the so--called ``hard'' 
Pomeron. Diagramatically, BFKL equation 
describes the $t-$channel exchange of ``gluonic ladder'' ( see Fig.1a )  
-- a concept familiar 
from the old--fashioned multi--peripheral model \cite{mpm}. 

At small momentum transfer, QCD perturbation theory is 
in general inapplicable, but one may choose to consider 
the scattering processes where the parton virtualities 
at the ends of the ladder are fixed to be large (for example, 
the scattering of two heavy quarkonium states \cite{AM1}).     
However, even then the partons can still ``diffuse'' to small 
values of transverse momenta toward the center of the ladder 
(diffusion in the log of transverse momenta  \cite{BFKL}), and at
sufficiently high 
energies the perturbative approach inevitably breaks down \cite{BARTELS}.
This argument 
was formulated rigorously by A.H. Mueller \cite{AM2}, 
who showed that the operator product expansion (which provides 
the basis of the perturbative approach) breaks 
down at high energies.
Another serious problem of the perturbative treatment has    
been made apparent by recent vigorous  calculations of the next--to--leading 
corrections to the BFKL equation \cite{NLO}. The NLO corrections 
appeared to be large, and drove the 
intercept of ``BFKL Pomeron'' significantly below the range of values 
suggested by phenomenology. 

Perturbative expansion of the scattering amplitude is possible only 
in the presence of a sufficiently large scale. 
As was mentioned above, at very high energies 
the external scale, which determines the parton virtualities at the ends 
of the ladder, becomes progressively unimportant, and the perturbative 
expansion loses justification. 
Therefore it looks plausible that the 
Pomeron is a genuinely non--perturbative phenomenon 
\cite{Nussinov}, \cite{VEN}.    
At present, non--perturbative phenomena can only be treated theoretically 
if they stem from relatively short distances, which requires the 
presence of a large scale.  
The main idea exploited in this letter 
is that such a scale exists in the QCD vacuum as a consequence 
of scale anomaly, and is related to the density of vacuum gluon fields due to 
semi--classical fluctuations; numerically, $M_0^2 \simeq 4 \div 6$ GeV$^2$ (see below). 
Because of the presence of this large scale, the perturbative expansion 
still makes sense; we are able also to evaluate explicitly the 
leading non--perturbative contribution due to the scale anomaly. 

There are two facts that support the feasibility of such an approach. First, 
the success of QCD sum rules is based on the use of a few first terms in the 
operator product expansion, which 
can be justified only if a sufficiently large scale associated 
with the vacuum structure exists \cite{NSVZ}. Second, 
the non--perturbative amplitude of low--energy dipole--dipole scattering was 
evaluated and found 
to be determined by the vacuum energy density, arising from the semi--classical 
fluctuations of gluon fields \cite{FK}.  
This latter example is encouraging, since the multi--peripheral model 
\cite{mpm} relates the amplitude of high--energy scattering to  
the low--energy interactions of partons.

Basing on these ideas, we propose an extension of the BFKL program to the 
non-perturbative domain; the 
key ingredient of our approach is the breakdown of scale invariance in QCD, 
reflected in scale anomaly. The concept of scale anomaly is rather 
general and was formulated 
long time ago \cite{JE},\cite{JE1},\cite{MS}; 
let us briefly recall its application to QCD.  
In the chiral limit of massless 
quarks, the Lagrangean of QCD is scale invariant on the classical, tree
level. 
This invariance is however broken by renormalization, which introduces a 
dimensionful scale once the interactions are switched on. This 
``dimensional transmutation'' phenomenon is fundamental for the 
understanding of scale dependence of the strong coupling constant \cite{asfr}, 
which is the basis of all applications of perturbative QCD.    
On the formal level, the breakdown of scale invariance in the theory 
is reflected by the non--conservation of scale current, and thus in 
the non--zero trace of the energy--momentum tensor $\theta^{\mu\nu}$ 
\cite{scale}. 
Scale anomaly leads to a set of powerful low-energy theorems 
for the correlation functions of gluon currents in the scalar channel 
\cite{NSVZ}.  

\begin{figure}[htbp] 
\begin{center}
\epsfig{file=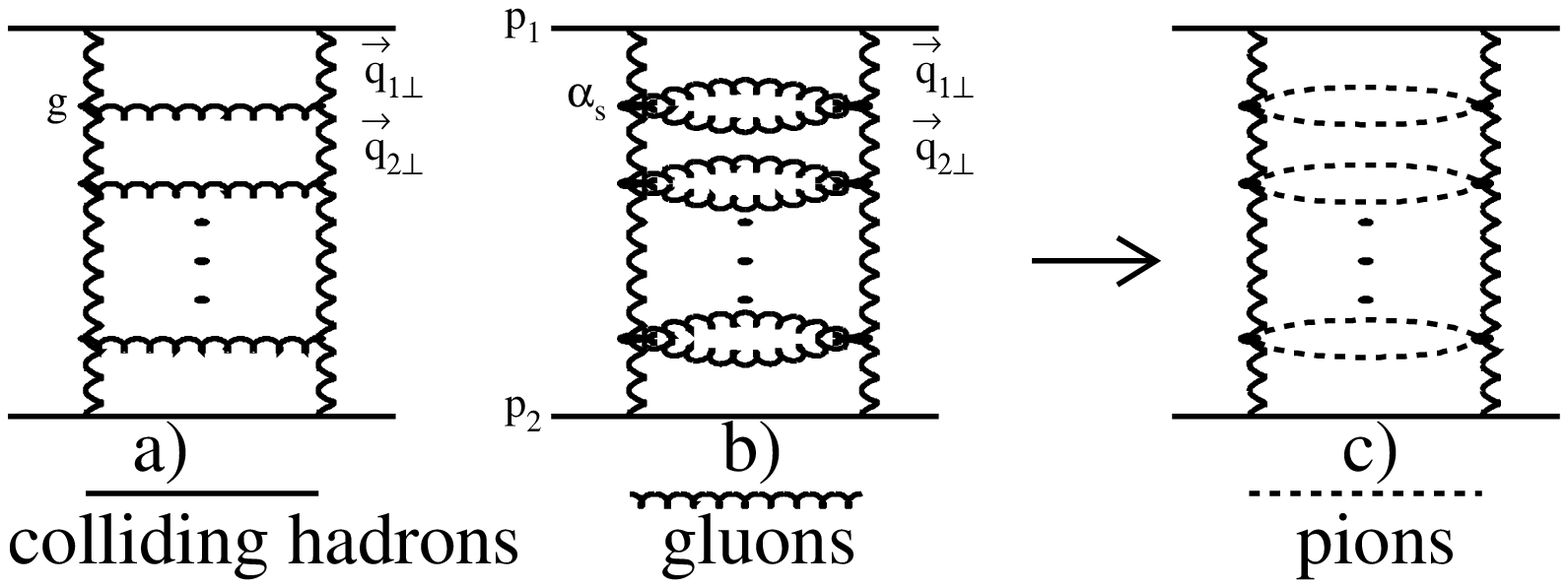,width=16cm}
\end{center}
\caption{\em Multi-peripheral ( ladder ) diagrams contributing to the leading--order BFKL 
(a) and ``soft" (b) and (c) Pomeron structure}
\label{fig1}
\end{figure}

The starting point of the approach that we propose in this letter 
is the following: 
among the higher order, $O(\alpha^2_S)$ ($\alpha_S= g^2/4\pi$) ,
corrections to the BFKL kernel we 
isolate a particular class of diagrams which include the propagation 
of two gluons in the scalar color singlet channel $J^{PC}=0^{++}$ 
(see Fig. \ref{fig1}-b). 
We will show that, as a consequence 
of scale anomaly, these, apparently  $O(\alpha^2_S)$, contributions become 
the {\it dominant} ones, $O(\alpha^0_S)$.

Indeed, let us consider the contribution of Fig. \ref{fig1}-b, which is one of 
the numerous corrections of the next--to--leading order to the
BFKL Pomeron. 
In perturbation theory, such corrections are $\propto
\alpha^2_S$; however, we note that if the two produced gluons in Fig.\ref{fig1}-b are in 
the scalar and colorless state, the vertex of their production, generated 
the four--gluon coupling in the QCD Lagrangean, is $\sim \alpha_s F^{\mu\nu\ a} 
F_{\mu\nu}^a$.  We observe that this vertex is therefore 
proportional to the trace of the
QCD energy--momentum tensor ( $\theta^{\mu}_{\mu} $) in the chiral limit 
of massless quarks:
\begin{equation}
\theta_\mu^\mu = 
\frac{\beta(g)}{2g} F^{a\alpha\beta} F_{\alpha\beta}^{a} \simeq 
- \frac{b g^2}{32 \pi^2} F^{a\alpha\beta} F_{\alpha\beta}^{a}; \label{trace} 
\end{equation} 
note that as a consequence of decoupling 
theorem \cite{svz2} the $\beta$ function in Eq.~(\ref{trace}) 
does not contain the contribution of heavy quarks ({\it i.e.} 
$b=\frac{1}{3}(11N-2N_f)=9$). 

The entire contribution of Fig.\ref{fig1}-b therefore 
appears proportional to the correlator of the QCD energy--momentum tensor. 
Let us now consider the spectral representation for this correlator:
\begin{equation}
\Pi(q^2) = i \int d^4 x\ e^{iqx}\ 
\bra{0} \T \left\{ \theta_\mu^\mu(x)\theta_\nu^\nu(0) \right\} \ket{0}
=
\int d\sigma^2 \frac{\rho_\theta (\sigma^2)}{\sigma^2-q^2-i\epsilon},
\label{corr}
\end{equation}
with the spectral density defined by
\begin{equation}
\rho_\theta (k^2) =
\sum_n (2\pi)^3 \delta^4(p_n-k)|\bra{n}\theta_\mu^\mu\ket{0} |^2 ,
\label{def_spect}
\end{equation}
where the phase-space integral is understood.
In lowest--order perturbation theory, the spectral density (\ref{def_spect})
is given by the contribution of two-gluon states; 
the calculation for $SU(N)$ color gives
\begin{eqnarray}
\rho^{\rm pt}_{\theta}(q^2) &=& 
\left ( \frac{bg^2}{32\pi^2} \right )^2\frac{N_c^2-1}{4\pi^2}\ q^4.
\label{pdens2}
\end{eqnarray}

However, at small invariant masses, perturbation 
theory inevitably breaks down; 
an important theorem \cite{NSVZ} for this correlator states that 
as a consequence of broken scale invariance of QCD,
\begin{equation}
\Pi(0)=-4\ \bra{0} \theta_\mu^\mu(0)\ket{0}.
\label{theorem}
\end{equation}

Since 
this theorem, as will become clear soon, is a corner--stone of our approach, 
let us briefly recall 
its proof \cite{NSVZ}. 
It is based on the fact that the expectation value of 
any operator $O$ of canonical dimension $d$ ($d=4$ for $\theta_\mu^\mu$) 
can be written down as 
\begin{equation}
\langle O \rangle \sim \left[M_0\ exp\left(-{\frac{8 \pi^2}{b g_0^2}}\right)\right]^d,
\label{renorm}
\end{equation}
where $g_0 \equiv g(M_0)$, and $M_0$ is the renormalization scale. 
On the other hand, the dependence of 
QCD Lagrangean on the coupling is $(-1/4 g_0^2)\ \tilde{F}^{a\alpha\beta} 
\tilde{F}_{\alpha\beta}^{a}$, where $\tilde{F} \equiv g F$ is the rescaled gluon 
field. By writing down the expectation value of the operator $O$ in the form 
of the functional integral, and by differentiating this expression with respect to $1/4 g_0^2$, 
one can therefore generate correlation functions of the operators 
$O$ and $\tilde{F}^2$. Differentiating once, 
one gets
\begin{equation}
i \int dx\ \langle\ \T \left\{ O(x)\ \tilde{F}^2(0) \right\} \rangle 
\equiv -{d \over d(1/4 g^2)} \langle O \rangle.
\label{correl}
\end{equation}
Combining (\ref{correl}) and (\ref{renorm}), and choosing $O(x) = \theta_\mu^\mu(x)$, 
furnishes the proof of the theorem (\ref{theorem}).  

Note that the r.h.s.\ of Eq.\ (\ref{theorem}) is divergent even in 
perturbation theory, and should therefore be regularized by
subtracting the perturbative part. The vacuum expectation value of
the $\theta_\mu^\mu$ operator then measures the energy density of 
non-perturbative
fluctuations in QCD vacuum, and the low-energy theorem (\ref{theorem})
 implies a sum rule for the spectral density:
\begin{equation}
\int \frac{d\sigma^2}{\sigma^2}
[\rho_\theta^{\rm phys}(\sigma^2)-\rho_\theta ^{\rm pt}(\sigma^2)]
=-4\ \bra{0}\theta_\mu^\mu(0)\ket{0}
=-16\ \epsilon_{\rm vac}
\ne 0 ,
\label{let}
\end{equation}
where the estimate for the vacuum energy density extracted from the 
sum rule analysis gives  
$\epsilon_{\rm vac}\simeq -(0.24 \mbox{ GeV})^4$ \ \cite{SVZ}.
In addition,  another sum rule \cite{SR,SVZ}, 
\begin{equation}
\int d\sigma^2 \rho_\theta^{\rm phys}(\sigma^2)
=
\int d\sigma^2 \rho_\theta^{\rm pt}(\sigma^2)
\label{duality}
\end{equation}
is implied by the quark--hadron duality.
Since the physical spectral density, $\rho_\theta^{\rm phys}$, should
approach the perturbative one,  $\rho_\theta^{\rm pt}$, at high
$\sigma^2$, the integral in Eq.\ (\ref{let}) is convergent.

According to (\ref{trace}) and (\ref{def_spect}), 
the l.h.s. of Eq.(\ref{let}) is apparently $O(g^4)$; however it is easy 
to see that this is not so by looking at the r.h.s. of this equation, 
which is renormalization group invariant, and does not depend on 
the coupling constant.  
This means that the l.h.s. must also be $O(g^0)$. Let us 
illustrate this formal argument by considering the spectral density 
(\ref{def_spect}) at small invariant 
mass \cite{VZ}. Small invariant masses imply small relative momenta 
for the produced particles, and at small momenta an accurate 
description of QCD is given by an effective chiral Lagrangean   
\begin{equation}
{\mathcal{L}} = {\frac{f_{\pi}^2}{4}}\ 
\tr\ \partial_{\mu} U \partial^{\mu}U^{\dagger}
 \ +\ \frac{1}{4}\ m_\pi^2 f_\pi^2 \ \tr \left(U + U^{\dagger} \right),
\label{nlsig}
\end{equation}
where $U = \exp\left( 2i \pi / f_{\pi} \right)$,  
$\pi \equiv \pi^a T^a$ and $T^a$ are the $SU(2)$ generators
normalized by $\tr\ T^a T^b = \frac{1}{2} \delta^{ab}$.
The trace of the energy--momentum tensor for this Lagrangian is 
(see, e.g., \cite{FK})
\begin{equation}
\theta_\mu^\mu =
 - 2\ \frac{f_{\pi}^2}{4}\ \tr\ \partial_{\mu} U \partial^{\mu} 
U^{\dagger} \ - \ m_\pi^2 f_\pi^2 \ \tr \left( U + U^{\dagger} 
\right). \label{trnl}
\end{equation}
Expanding this expression (\ref{trnl}) in powers of the pion field,
one obtains, to the lowest order, 
\begin{equation}
\theta_\mu^\mu =
-\partial_\mu \pi^a \partial^\mu\pi^a +2 m_\pi^2 \pi^a \pi^a + \cdots ,
\label{trl}
\end{equation}
and this leads to an elegant result \cite{VZ} in the chiral limit of
vanishing pion mass: 
\begin{equation}
\bra{\pi^+\pi^-} \theta_\mu^\mu \ket {0} = q^2 \ .
\label{vz}
\end{equation}
This result for the
coupling of the operator $\theta_\mu^\mu$ to two pions can be
immediately generalized for any (even) number of pions using
Eq.~(\ref{trnl}). 
The expression (\ref{vz}) is manifestly $\sim O(g^0)$, and shows 
that the spectral density of the scalar gluon operator $\sim g^2 F^2$ 
is independent of the coupling constant $g$ as a consequence 
of scale anomaly. While we have used an effective chiral Lagrangean 
to illustrate how the dependence on the coupling constant gets 
``eaten'' by the scale anomaly, this phenomenon is very general and 
does not depend on the specific model for the spectral density. 
One way of understanding the disappearance of the coupling constant 
in the spectral density of the $g^2 F^2$ operator is to assume that the 
non-perturbative QCD vacuum is dominated by the semi--classical fluctuations 
of the  gluon field. Since the strength of the classical gluon field 
is inversely proportional to the coupling, $F \sim 1/g$, 
the quark zero modes, and the spectral density of their pionic excitations, 
appear independent of the coupling constant. 

Armed with this knowledge, we are ready to see that   
 the contribution to the 
next--to--leading order BFKL kernel that describes the production of 
two gluons in the color singlet, scalar state, which is formally 
$\sim O (g^4)$, as a consequence of scale anomaly can become 
the leading one, $\sim O(g^0)$.    
(Of course, the perturbative part of this contribution is
still $\sim O (g^4)$ and has been taken into account in 
the next-to-leading order
BFKL Pomeron).  
Therefore, we want to build a multi-peripheral model for the ``soft"
Pomeron in which  hadrons are produced (mostly two pions, see
Refs.\cite{FK}, \cite{EFK}) due to  
exchange of two gluons in the $t$--channel (see Fig.\ref{fig1}-c ).
The only dimensional scale in this approach appears in $\rho^{phys}$ and
can be estimated directly from sum rules of \eq{let}. It turns out that the
characteristic mass ($M^2_0$) in \eq{let} is rather large \cite{FK}, \cite{EFK}  $M^2_0
\,\approx\,4\,$GeV$^2$ (the original analysis of \cite{NSVZ}, \cite{MS1} yielded even 
somewhat bigger value $M^2_0
\,\approx\,6\,$GeV$^2$). This is the largest scale which exists in non--perturbative QCD 
\cite{NSVZ}, \cite{Shuryak}. In the framework of the instanton approach, the large 
magnitude of $M_0$ was shown to be a consequence of strong color field inside 
the instanton \cite{Shuryak}.

This value determines the scale of all dimensional
parameters of the Pomeron trajectory as well as the typical transverse
momentum of produced particle in the Pomeron. It is interesting to notice
that the experimental value for the slope $\alpha'_P(0)$ of the Pomeron trajectory
(in the standard notation, $\alpha_P(t) = 1 + \Delta + \alpha'_P(0)\, t $)  $\alpha'_P(0) =
0.25\,$GeV$^{-2}$\cite{DL} is very close to $1/M^2_0$.

We start our calculation with diagrams of Fig.\ref{fig1}-b 
in the leading log $s$ approximation of pQCD where we sum only contributions
of the order of  $( \alpha_S\,\ln s )^n$. In this approximation 
the propagators of the t-channel gluons can be written in a simple form
\cite{GLR}
\beq \label{4}
G_{\mu \nu}(q^2_i) \,\,=\,\,\frac{g_{\mu \nu }}{q_i^2}
\,=\,\frac{1}{q^2_{i,\perp}}\,\times\,\frac{2\,q_{i,\perp,\mu}\,
q_{i,\perp,\nu}}{\alpha_i \,\beta_i \,s}\,\,+\,\,O \left(\frac{1}{s}
\right) \,\,,
\eeq
where we use the Sudakov decomposition for momenta $q_i$ along the
momenta of colliding particles ( $p_1$ and $p_2$ in Fig.\ref{fig1}-b ),
namely, 
\beq \label{5}
q_{i,\mu}\,\,=\,\,\alpha_i\,p_{1,\mu}\,\,+
\,\,\beta_i\,p_{2,\mu}\,\,+\,\,q_{i,\perp,\mu}\,\,;
\eeq
\eq{4} corresponds to Weizs{\"a}cker--Williams approximation 
for the gluon field of a fast--moving hadron. 

The ladder diagram of Fig.\ref{fig1}-b for emission of $n$-pairs is equal
to
\begin{eqnarray} 
\sigma_n(Q^2)\,&=&  \as^2 \int
\Gamma_{\mu} \Gamma_{\nu}   
 \prod_{i=1}^{i=n+1} 
\frac{ s\,\,d \alpha_i d \beta_i d^2 q_{i,\perp}}{2 ( 2
\pi)^3}\,\delta( \alpha_{i}\beta_{i + 1} s - M^2_{i,\perp}) \,\Phi_i \label{6}\\
& \cdot&
\Gamma_{\mu_{i},\mu_{i + 1}}\Gamma_{\nu_{i},\nu_{i +
1}} G_{\mu_i,\mu_{i + 1} }(q^2_{i})G_{\nu_i,\nu_{i + 1}
}((Q - q_i)^2)\,\Gamma_{\mu_i+1}\Gamma_{\nu_i+1}\,\,\,,\nonumber
\end{eqnarray}
where $\sigma(Q^2=0)$ is the total cross section of $n$-pairs production
($ \sigma_{n}$) and $\Phi_i$ is the phase space factor for two identical
particles with total mass $M_i$ which is equal to
\beq \label{7}
\Phi_i\,\,=\,\,\frac{1}{2}\int \,\frac{d^3 k_1}{(2 \pi)^3 2 \omega_1 2
\omega_2}\,\delta( \omega_1 + \omega_2 - M_i )\,\,=\,\,\frac{1}{32
\pi^2}\,\,.
\eeq
In \eq{6} $\Gamma_{\mu_{i},\mu_{i + 1}}$ is the vertex of gluon pair
production. It is easy to calculate that it is equal to
\beq \label{8}
q_{i,\perp,\mu_i}\,q_{i,\perp,\mu_{i+1}}\,\Gamma_{\mu_{i},\mu_{i + 1}}
\,\,=\,\,3\,\vec{q}_{i,\perp}\cdot\vec{q}_{i+1,\perp}\,\,,
\eeq
after projecting on the colorless state with $J^{PC}=0^{++}$.

Using the kinematic relation $\alpha_{i} \,\beta_{i + 1} \,s = M^2_i
\,+\,k^2_{i,\perp}$, where $\vec{k}_{i,\perp}\,\,=\,\,\vec{q}_{i,\perp}
\,-\,\vec{q}_{i+1,\perp}$ and performing integration over $\alpha_i$
explicitly,  we can rewrite \eq{6} in the simple form
\beq \label{9}
\sigma_n(Q^2)\,\,=\,\,\as^2 \,\frac{\left(  \ln s \right)^n}{n!}
 \prod^{i = n+1}_{i = 1}\,\,\frac{4 \cdot  9 \cdot \as^2}{32 \pi^2}\,
\frac{
( 
\vec{Q}_{\perp} 
-\vec{q}_{i,\perp}) 
\cdot(
\vec{Q}_{\perp} 
-\vec{q}_{i+1,\perp})}{q^2_{i,\perp} \,( \vec{Q} - \vec{q}_i )^2_{\perp} }
\,\,\frac{d M^2_i}{\left( M^2_i + k^2_{i,\perp} \right)^2}\,\,.
\eeq
For forward scattering $Q^2_{\perp} = 0$ and \eq{9} leads to  
power--like
behavior of the total cross section:
\beq \label{10}
\sigma_{tot} = \sum^{\infty}_{n = 0}\, \sigma_n \, =\,\sigma^{BORN} s^{\Delta}
\,\,,
\eeq
where 
\beq \label{11}
\Delta \,\,=\,\,\frac{\as^2\ 18}{32 \pi^2}\int \frac{d k^2_{\perp}\,\,d
M^2}{ \left( M^2
\,+\,k^2_{\perp}\right)^2}\,\,.
\eeq
and $\sigma^{BORN}$ is the cross section due to two gluon exchange
\beq \label{BORN}
\sigma^{BORN}\,\,=\,\,\as^2\,\int\,d^2 q \Gamma_{\mu} \Gamma_{\nu}
G_{\mu, \mu_1}(q^2_{\perp})\,\,G_{\nu,
\nu_1}(q^2_{\perp})\,\,\Gamma_{\mu_1}
\Gamma_{\nu_1}\,\,.
\eeq

\eq{11} can be easily rewritten through the perturbative spectral density 
$\rho^{pQCD}_{\theta}$ that was evaluated above (see Eq. (\ref{pdens2})):
\beq \label{13}
\Delta \,\,=\,\frac{\pi^2}{2} \times \left( \frac{8 \pi}{b}
\right)^2\,\times \frac{18}{32 \pi^2}\,\int\,\frac{ d M^2}{M^6}
\,\rho^{pQCD}_{\theta}( M^2 )\,\,,
\eeq
where $\as(M^2) = 4 \pi/(b \ln(M^2/\Lambda^2))$.

Our main idea is to separate non--perturbative and perturbative 
contributions to the spectral density of the scalar gluon operator by 
replacing $\rho^{pQCD}_{\theta}( M^2 )$ with 
$(\rho^{phys}_{\theta}( M^2 )\,\,- \,\,\rho^{pQCD}_{\theta}( M^2 )) + 
\,\,\rho^{pQCD}_{\theta}( M^2 )$. The purely perturbative contribution 
is of the order of $O(g^4)$, and has been evaluated before \cite{NLO}.
For the non--perturbative contribution, in which we are interested here, 
we have
\beq \label{14}
\Delta \,\,=\,\frac{\pi^2}{2} \times \left( \frac{8 \pi}{b}
\right)^2\,\times \frac{18}{32 \pi^2}\,\int\,\frac{ d M^2}{M^6}
\,\left(\,\rho^{phys}_{\theta}( M^2 )\,\,-\,\,\rho^{pQCD}_{\theta}( M^2
)\,\right)\,\,.
\eeq
To estimate the integral in \eq{14} we will use the chiral  
approach to $\rho^{phys}_{\theta}$ described above, 
namely,
\beq \label{15}
\rho^{phys}_{\theta}(M^2) \,  = \,\,
\frac{3}{32\pi^2}\,M^4\,\,,
\eeq
which corresponds to diagram of Fig. \ref{fig1}-c \footnote{Of course, 
the chiral approach cannot be extrapolated up to $M_0 \simeq 2\ \rm{GeV}$; 
at large $M$ the spectral density (\ref{15}) should be corrected by a 
phenomenological form-factor expressed in terms of experimental $\pi \pi$
phase shifts \cite{FK}; however, 
since our sensitivity to the large $M$ region is only logarithmic (see (\ref{14})), 
in this paper we will use the simplified ansatz (\ref{15}).}.

It is instructive to establish a qualitative relation between the matching 
parameter $M_0$ and the energy density of QCD vacuum using the spectral 
density (\ref{15}) and the sum rule (\ref{let}).  Since perturbative spectral 
density $\rho^{pQCD}_{\theta}( M^2 )$ at moderate $M$ is much smaller than 
$\rho^{phys}_{\theta}(M^2)$, Eqs (\ref{15}) and (\ref{let}) lead to 
the following approximate relation \cite{MS1}:
\beq
M_0^2 \ \ \simeq \ \ 32 \pi \left\{ {|\epsilon_{\rm vac}| \over N_f^2 - 1}\right\}^{\frac{1}
{2}}, \label{rel}
\eeq
which shows that the matching scale $M_0$ is directly determined by the 
energy density of the vacuum. Since $\epsilon_{\rm vac} \sim N_c^2$, the magnitude of 
$M_0$ is proportional to $N_c^2/N_f^2$.  

Collecting all numerical 
factors and substituting $b = 9 $, we obtain
\beq \label{16}
\Delta\,\,=\,\, \frac{1}{48}\,\,\ln \,\frac{M^2_0}{4 m^2_{\pi}}\,\,.
\eeq
Let us discuss the result of these simple calculations:
\begin{enumerate}
\item\,\, \eq{10} and \eq{16} say that our approach leads to the exchange of
the Pomeron with the
intercept $\alpha(0) = 1 + \Delta > 1$.  To the best of our knowledge,  
this is the only approach in the framework of QCD which leads to a 
``soft'' Pomeron;

\item\,\,\eq{10} is a consequence of a direct generalization 
of the BFKL approach to the non--perturbative domain.
It
should be recalled that the only theory where the Pomeron naturally
appears is the BFKL Pomeron in 2\,+\,1 dimensional QCD
\cite{2QCD};

\item\,\, 
\eq{16} gives $\Delta \,=\,0.082$ for $M^2_0 = 4 \,$GeV$^2$ \cite{FK}  in
good agreement with the
phenomenological intercept of the ``soft" Pomeron, $ \Delta = 0.08$
\cite{DL}; it should be noted however that the precise value of the 
matching scale $M^2_0$ as extracted from the low--energy theorem (\ref{let}) 
depends somewhat on detailed form of the spectral density, 
and can vary within the range of $M^2_0 = 4 \div 
6\,$GeV$^2$ \cite{NSVZ}, \cite{FK}. 
Fortunately, the dependence of \eq{16} on $M_0$ is only 
logarithmic, and varying it in this range leads to 
\beq
\Delta = 0.08 \div 0.1. \label{number} 
\eeq

\item\,\, As we have already stressed, in our approach the only 
dimensionful parameter is $M^2_0$; its large value implies the dominance of rather 
short distances in the ``soft'' Pomeron structure. This fact is in 
agreement with a number of experimental and phenomenological observations:
\begin{itemize}
\item\,\, The value of the slope for the ``soft" Pomeron trajectory
$\alpha'_P(0) \,=\,0.25\,$GeV$^{-2}\,\,\ll\,\,\alpha'_R(0) = 1\,$GeV$^{-2}$,
where $\alpha'_R$ is the slope of the Reggeon trajectory;
\item\,\, The experimental slope of the diffraction production of the
hadron system with large mass is approximately two times smaller the slope
for the elastic scattering. It means that the proper size of the triple
Pomeron vertex is rather small. For our Pomeron it should be on the order of 
$1/M^2_0 \approx 0.25 \,$GeV$^{-2}\,\ll\,B_{el} = 10\,$GeV$^{-2}$;
\item\,\, The HERA data \cite{HERA} on diffractive $J/\Psi$  production
in DIS show that the $t$ - slope for elastic diffractive dissociation ($
\gamma^* + p \rightarrow J/\Psi + p  $)
is larger than the $t$ - slope for
the inelastic one ($\gamma^* + p \rightarrow J/\Psi + X  $, where $X$
is a high--mass hadronic system). This shows the existence of two different 
scales in the proton, one of which is determined by its size, and another 
one by the correlation length $\sim 1/M_0$ of the gluon field inside.
\end{itemize}
\item\,\, Non--trivial azimuthal dependence observed recently in diffractive 
production of scalar mesons \cite{wa102} can be explained \cite{EK} if one 
adopts the idea that the effective coupling of the Pomeron to mesons is dictated 
by scale anomaly.   
\end{enumerate}

The disappearance of the dependence on the coupling constant, which is the central point of  
our approach, may seem puzzling. However let us mention again that  
this result can be easily 
understood if we recall 
the interpretation of the non--zero v.e.v. of the trace of the 
energy--momentum tensor as being due to the semi--classical fluctuations of 
gluon fields. Since the strength of the classical gluon field is $F^2 \sim 1/\alpha_S$, 
quark zero modes, and their pionic excitations, 
appear independent of the coupling, $\sim O(\alpha_S^0)$. 
We therefore envision the Pomeron as a $t-$channel exchange of two gluons, which 
scatter off semi--classical fluctuations of vacuum gluon fields; this scattering 
is accompanied by the excitation of quark zero modes in the vacuum, resulting 
in the production of pions. Amazingly similar picture of the ``soft'' Pomeron 
was anticipated by Bjorken \cite{Bj}. These effects also manifest themselves 
in the low--energy scattering of heavy quarkonia; 
the magnitude of the scattering amplitude was found \cite{FK} to be 
determined by the energy density of the non--perturbative QCD vacuum.

Let us discuss the dependence of our result on the numbers of colors, $N_c$, and 
flavors, $N_f$. Two limits are of theoretical interest: {\it i)} $N_c \to \infty$, 
$N_f, \ g^2\ N_c$ fixed; {\it ii)} $N_c \to \infty$, $N_f/N_c, \ g^2\ N_c$ fixed. 
The case {\it i)} corresponds to the large $N_c$ limit proposed by 't Hooft \cite{Hooft}, 
while {\it ii)} is the basis of ``topological expansion'' suggested by Veneziano \cite{VEN}. 

\begin{figure}[htbp]
\begin{center}
\begin{tabular}{c c}
\epsfig{file=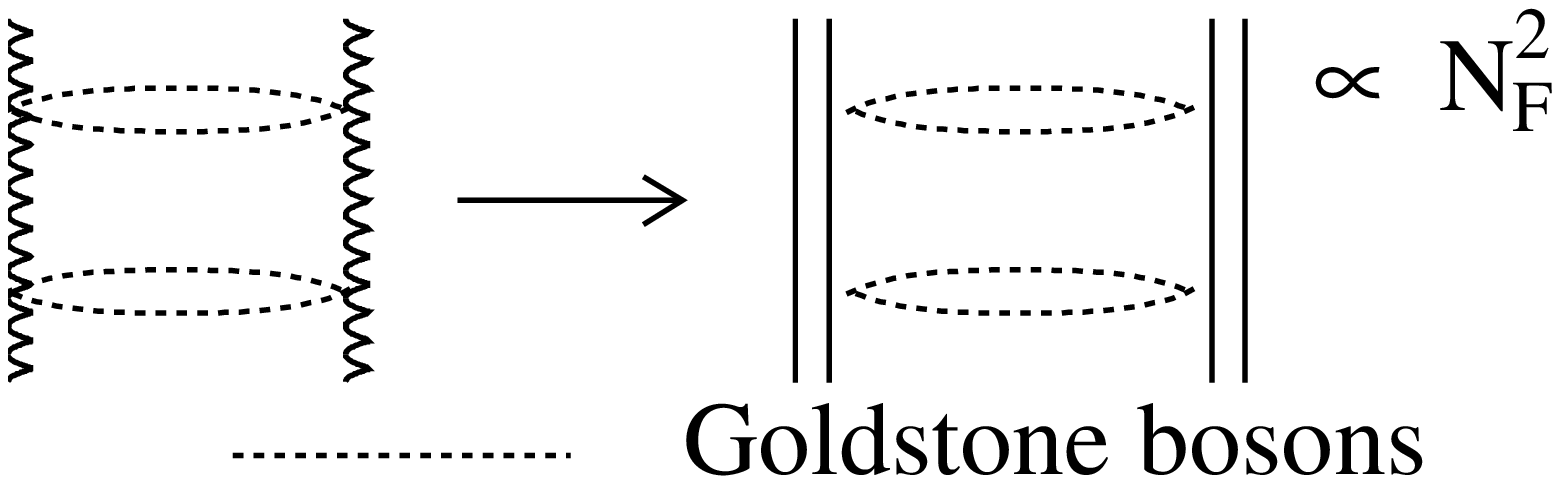,width=8cm,height=4cm}&
\epsfig{file=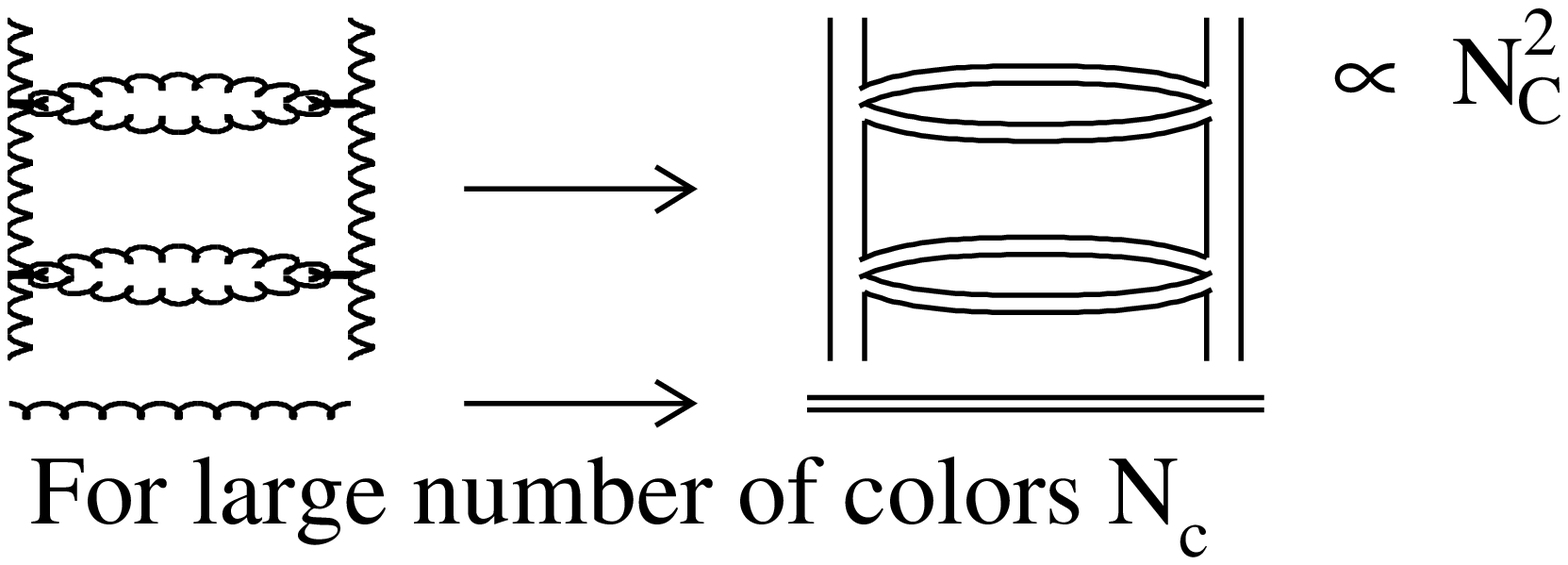,width=8cm,height=4cm}\\
Fig. 2-a & Fig.2-b
\end{tabular}
\end{center}
\caption{\em A simple illustration of the appearance of the $N^2_f/N^2_c$
factor in the Pomeron intercept.}
\label{fig2}
\end{figure}

Since the number of Goldstone bosons contributing to the non--perturbative spectral 
density (\ref{15}) for spontaneously broken $SU_L(N_f) \times SU_R(N_f)$ is 
equal to $N_f^2 -1$, it is evident from \eq{14} that $\Delta \sim N^2_f/N^2_c$ 
(note that \eq{14} contains $b = 1/3 (11 N_c - 2 N_f)$ in the denumerator).      
A simple graphic illustration of this dependence is given in
Fig.\ref{fig2}. 
Therefore, our approach 
may be considered as a realization of general
ideas, proposed long time ago by Veneziano \cite{VEN}, 
that the ``soft" Pomeron should be found keeping $N^2_f/N^2_c$ fixed.

Let us discuss now the large $N_c$ limit {\it i)}, which corresponds to pure 
gluodynamics. Naively, since $\Delta$ was found proportional to $N^2_f/N^2_c$, 
one may conclude that in this limit $\Delta$ vanishes, and the cross section 
does not grow with energy. This conclusion is, however, immature. Indeed, 
the physical spectrum in the scalar channel in gluodynamics contains a scalar glueball
\footnote{We thank S. Nussinov and E. Shuryak for stressing the r{\^{o}}le of the scalar 
glueball for our approach in the case of pure gluodynamics.}; 
its couplings to mesons are suppressed by $1/N_c$ (see, e.g., \cite{Col}), 
so it should be very narrow. Therefore, the spectral density \eq{15} should  
be replaced by 
  
\beq \label{specN}
\rho^{phys}_{\theta}(M^2) \,  = \,\,R\, M_R^6\ \delta(M^2 - M_R^2) + pert.\ contribution 
\eeq
where $M_R$ is the scalar glueball mass, and $R$ is its residue; the factor $M_R^6$ 
is introduced to make $R$ dimensionless. 
Using \eq{specN} in the sum rule (\ref{let}), we get a simple relation
\beq \label{residue}
R = 16\ {\frac{|\epsilon_{{\rm vac}}|}{M_R^4}}.
\eeq
With \eq{specN} and \eq{residue}, \eq{14} becomes
\beq
\Delta = \frac{288 \pi^2}{b^2}\ \frac{|\epsilon_{{\rm vac}}|}{M_R^4}; \label{interN}
\eeq
since $M_R \sim N_c^0$, $\epsilon_{{\rm vac}} \sim N_c^2$, and $b \sim N_c$, 
\eq{interN} is well--defined in the large $N_c$ limit. 

One could try to estimate the value of $\Delta$ given by \eq{interN} for $N_c=3$; 
this requires the knowledge of the mass of the scalar glueball in pure gluodynamics. 
Recent lattice result \cite{VW} gives $M_R \simeq 1.65$ GeV; assuming that the main 
contribution to the energy density of the vacuum is due to gluons and using, as before, 
$\epsilon_{{\rm vac}} \simeq (0.24\ {\rm GeV})^4$, $b = 11 N_c/3 = 11$, we get the value 
\beq
\Delta_{gluodynamics} \simeq 0.01,
\eeq
which is significantly smaller than our result (\ref{number}) for the world with light quarks. 
This indicates that the presence of light quarks in the theory leads to a much faster 
growth of the cross section with energy. 

The key question is whether one can prove the theoretical self-consistency of our approach.
Indeed, our classification of the contributions to the scattering amplitude is still 
based on the expansion in powers of $\alpha_S$, in which we have 
isolated the term $\sim O(\alpha_S^0)$ emerging as a consequence of scale anomaly. 
This term is the leading one only if the coupling constant $\alpha_S$ is sufficiently 
small. The magnitude of the coupling depends on the renormalization scale $M_0$. 
Since this dimensionful scale extracted from the sum rule analysis
is large, $M_0 = 4 \div 6$ GeV$^2$,      
the coupling constant indeed appears to be small, $\alpha_S(M_0^2) \ll 1$. 
This fact insures that perturbative corrections 
to the kernel are smaller than the leading, $\sim O(\alpha_S^0)$, term, and 
should be taken into account in the framework of conventional BFKL approach. 
Our approach yields a natural rapidity scale 
$(Y_0 = ln (M_0^2 / 4 m_{\pi}^2) \simeq 2 \div 3)$ 
for BFKL kernel above which the perturbative approach can be applied. It is interesting to 
note that for $Y \geq Y_0$ the next--to--leading order corrections are well under control 
\cite{Lipatov}; however the interplay between ``soft''  and ``hard''  
physics still has to be understood. 

We therefore believe that our proposal can lead to a systematic 
theoretical approach 
to the Pomeron in QCD. 
Of course, one cannot exclude {\it a priori} a different view of 
the Pomeron structure as coming from large distances, $R \gg 1/M_0$. 
However, experimental data (see point 4 of our discussion above) 
suggest that the phenomenological Pomeron indeed originates at 
small distances $R \sim \sqrt{\alpha'_P(0)} \sim 1/M_0$. 

Let us discuss the relation of our approach to other existing approaches to soft scattering. 
Very similar ideas of the dominance of semi--classical vacuum 
gluon fields were developed in 
Refs. \cite{msv}. Our approach is complementary to these ideas, giving a 
natural explanation of the energy behavior of the soft scattering amplitude, which 
previously had to be taken phenomenologically\footnote{Very recently, 
a new attempt to describe the energy dependence has been made in Ref. \cite{Kai}, 
with a different result.}. Let is note also that the correlation 
length of gluon fields, which was taken from the lattice QCD calculations in Refs 
\cite{msv}, in our approach appears only implicitly and is determined 
from the analysis of low--energy theorems.

The dominance of classical gluon field configurations in high--energy collisions 
is the key idea of the approach proposed by McLerran and Venugopalan \cite{MV} 
and developed in Refs. \cite{MV1}. In this approach, 
the r\^{o}le of dimensionful parameter is played by the density of color charges in the 
transverse plane, rather than by the vacuum energy density. 
 In our opinion, this is a plausible assumption at very 
high energies and/or for sufficiently heavy nuclei for low partial amplitudes (central region 
in the impact parameter plane). Since we focus our attention on the behavior of the total 
cross section, which is determined by large distances in the impact parameter space, 
and therefore small density of the color charge, the relevance of the scale $M_0^2$ associated 
with the vacuum field strength should not be surprising.  
We feel that the approach of \cite{MV}, \cite{MV1} can describe the inclusive 
cross section, while ours is suited for the description of the total cross section. 
Indeed, the multiplicity associated with our multi--peripheral ladder 
$\sim 2\ \alpha(0)\ ln\ s$ 
is rather small compared to the expectations of \cite{MV}, \cite{MV1}.  
It would be extremely interesting to understand better 
the relationship between the two approaches.  
\vskip0.3cm

{\large\bf Acknowledgements}
\vskip0.3cm

We thank Ian Balitsky, James Bjorken, Wilfried Buchm\"{u}ller, G\"{u}nter Dosch, 
John Ellis, Hirotsugu Fujii, Errol Gotsman, Bob Jaffe, Alexei Kaidalov, T.D. Lee, Uri Maor, 
{\mbox{Larry McLerran}}, 
{\mbox{Al Mueller}}, Shmuel Nussinov, Rob Pisarski, Misha Ryskin, Edward Shuryak, 
Chung--I Tan, Larry Trueman and 
Raju Venugopalan for very fruitful discussions of problems related to this work.

The work of D.K. was supported by the US Department of Energy 
(Contract \# DE-AC02-98CH10886) and 
RIKEN. 
The research of E.L. was supported in part by the Israel Science 
Foundation, founded by the Israeli Academy of Science and Humanities, 
and BSF \# 9800276.

\end{document}